\documentclass[%
 reprint,
superscriptaddress, 
showkeys,
 amsmath,amssymb,
 aps,
]{revtex4-1}

\usepackage{graphicx}
\usepackage{dcolumn}
\usepackage{bm}

\usepackage{gensymb}
\usepackage{amssymb}

\usepackage[utf8]{inputenc}
\usepackage[T1]{fontenc}

\begin{document}

\preprint{APS/123-QED}

\title{Effect of Low-Damage Inductively Coupled Plasma on Shallow NV Centers in Diamond}

\author{Felipe Fávaro de Oliveira}%
\affiliation{$3.$ Institute of Physics, Research Center SCoPE and IQST, University of Stuttgart, 70569 Stuttgart, Germany}
\author{S. Ali Momenzadeh}%
\affiliation{$3.$ Institute of Physics, Research Center SCoPE and IQST, University of Stuttgart, 70569 Stuttgart, Germany}
\author{Ya Wang}%
\affiliation{$3.$ Institute of Physics, Research Center SCoPE and IQST, University of Stuttgart, 70569 Stuttgart, Germany}
\author{Mitsuharu Konuma}
\affiliation{Max Planck Institute for Solid State Research, 70569 Stuttgart, Germany}
\author{Matthew Markham}
\affiliation{Element Six Innovation, Harwell Oxford, Didcot, Oxfordshire, OX11 0QR, United Kingdom}
\author{Andrew M. Edmonds}
\affiliation{Element Six Innovation, Harwell Oxford, Didcot, Oxfordshire, OX11 0QR, United Kingdom}
\author{Andrej Denisenko}%
\email{Corresponding author: a.denisenko@physik.uni-stuttgart.de}
\affiliation{$3.$ Institute of Physics, Research Center SCoPE and IQST, University of Stuttgart, 70569 Stuttgart, Germany}
\author{Jörg Wrachtrup}%
\affiliation{$3.$ Institute of Physics, Research Center SCoPE and IQST, University of Stuttgart, 70569 Stuttgart, Germany}
\affiliation{Max Planck Institute for Solid State Research, 70569 Stuttgart, Germany}

\date{\today}

\begin{abstract}
Near-surface nitrogen-vacancy ({NV}) centers in diamond have been successfully employed as atomic-sized magnetic field sensors for external spins over the last years. A key challenge is still to develop a method to bring NV centers at nanometer proximity to the diamond surface while preserving their optical and spin properties. To that aim we present a method of controlled diamond etching with nanometric precision using an oxygen inductively coupled plasma (ICP) process. Importantly, no traces of plasma-induced damages to the etched surface could be detected by X-ray photoelectron spectroscopy (XPS) and confocal photoluminescence microscopy techniques. In addition, by profiling the depth of NV centers created by $5.0$ keV of nitrogen implantation energy, no plasma-induced quenching in their fluorescence could be observed. Moreover, the developed etching process allowed even the channeling tail in their depth distribution to be resolved. Furthermore, treating a ${}^{12}$C isotopically purified diamond revealed a threefold increase in T${}_2$ times for NV centers with $<4$ nm of depth (measured by NMR signal from protons at the diamond surface) in comparison to the initial oxygen-terminated surface.
\end{abstract}

\maketitle

 

The negatively-charged nitrogen-vacancy ({NV}) center in diamond has attracted increasing attention due to its outstanding properties. It is an atomic-sized, bright and stable single photon source\cite{reviewJWrachtrup} with relatively long coherence times, ranging milliseconds in isotopically purified single crystal diamond layers\cite{JahnkeC12,GopanTHz}. Additionally, its electron spin can be coherently manipulated by microwave and readout optically at room temperature. In the recent years the use of near-surface (shallow) NV centers as sensors to detect external nuclear\cite{staudacher,MaminNMR2013,NMRMueller} and electronic\cite{Bernhard,singleProteinFazhan} spins have been successfully demonstrated. However, since the signal detection relies on the relatively weak dipolar coupling to the targeted external spins, decaying proportional to r${}^{-3}$ (with $r$ being the distance between the targeted and sensor spins), NV centers must be located close to the diamond surface ($< 5$ nm)\cite{staudacher,MaminNMR2013}. 

Up to now, the engineering of near-surface NV centers has relied mostly on low-energy nitrogen implantation\cite{pezzagnaYield} or epitaxial growth of high quality nitrogen-doped CVD diamond followed by electron\cite{dGrownT2_KOhno} or ion irradiation\cite{C12irr}. Furthermore, NV centers can be brought closer to the diamond surface by post-treatments such as thermal oxidation\cite{LorentzAB, OxPlasmaKim} and etching in plasma\cite{OxPlasmaKim,VertDist,NLplasma}.

\begin{figure}[t]
\includegraphics[width=\columnwidth]{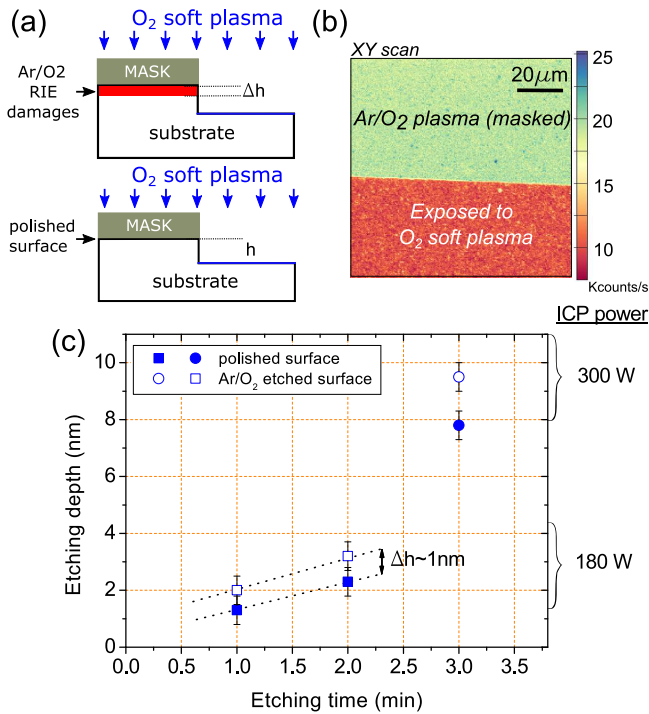}
\caption{\label{ICPfigure} (a): schematic representation of the performed etching experiment. (b): {XY} confocal photoluminescence scan from the diamond surface treated with Ar/O${}_2$ RIE plasma; the fluorescence contrast between the masked- and oxygen soft plasma-treated regions can be clearly seen. (c): etching depth measurements by AFM from oxygen soft plasma-treated polished and previously Ar/O${}_2$ RIE-etched surfaces; a difference of $1-2$ nm can be seen (represented by $\Delta$h). The etching rate for 180 W of ICP power was $1.5 \pm 1.0$ nm/min.}
\end{figure}

A major drawback for thermal oxidation is the uncertainty in the resulting etching rate and infeasibility of selective etching. Overcoming these issues, plasma processes are widely employed, providing a smooth and uniform method to selectively etch diamond. In particular for reactive ion etching (RIE) processes, the presence of a bias between the plasma source and the sample leads to ion bombardment on the diamond surface. This results in an enhanced etching rate and directionality of the process, but also produces a highly damaged layer containing vacancies and implanted ions of a few nanometers below the diamond surface\cite{PlasmaDamXPS}. If located in the vicinity of NV centers, these defects can lead to suppression of the photoluminescence emission and degraded spin properties\cite{OxPlasmaKim,NLplasma}. Thus a precise, uniform and high-selective etching process, by which the damages to the etched diamond surface are avoided, is highly desired. To that aim we present a low-pressure oxygen inductively coupled plasma (ICP) process with nanometric etching precision and high reproducibility. All reported plasma-related processes were performed in an Oxford PlasmaPro NGP80 machine equipped with an additional ICP source (Oxford Plasma Technologies). The effect of the developed etching process in the optical and spin properties of near-surface NV centers in diamond CVD layers will be explored in details. 

The optimized plasma recipe consists of two steps: first the RIE source with $10$ mTorr constant chamber pressure and $30$ W of power is used to ignite the plasma for a few seconds. Afterward the RIE source is switched off and the plasma is sustained only by the ICP source, which is located at a remote position from the substrate holder in the chamber. Under such conditions the sample surface is exposed mainly to neutral chemical radicals. This is supported by the fact that using both sources simultaneously showed no significant change in the DC bias, while varying the ICP power up to $300$ W and keeping the RIE power constant at $30$ W. This plasma process will be referred further as ``oxygen soft plasma'' in this paper.       

A high-purity single-crystal $[100]$-oriented electronic grade CVD diamond with an as-polished surface was taken as a substrate. The initial root mean squared (RMS) surface roughness was measured to be in the range of 1 nm by atomic force microscopy (AFM), allowing a detailed analysis in the quality of the post-treated surface. A schematic representation of the experiment is presented in figure \ref{ICPfigure}(a). The sample was masked by lithography-patterned AZ $5214$ E (MicroChemicals) photoresist to protect part of the polished surface. Subsequently, the sample was exposed to Ar/O${}_2$ RIE plasma for 16 minutes ($100$ sccm and $11$ sccm respectively, $37.5$ mTorr chamber pressure and $70$ W RIE power). Next, the polished and the Ar/O${}_2$ RIE etched regions were again patterned by optical lithography, leaving stripes exposed to the subsequent treatment with the oxygen soft plasma. Such procedure yields four different areas on the diamond surface, namely (I.) as-polished, (II.) polished combined with the oxygen soft plasma, (III.) Ar/O${}_2$ RIE as-etched and (IV.) Ar/O${}_2$ RIE etched combined with the oxygen soft plasma. The sample was cleaned by boiling in a triacid mixture of H${}_2$SO${}_4$:HNO${}_3$:HClO${}_4$, $1$:$1$:$1$ volumetric ratio for three hours, referred as wet chemical oxidation (WCO). 

\begin{figure}[t]
\includegraphics[width=\columnwidth]{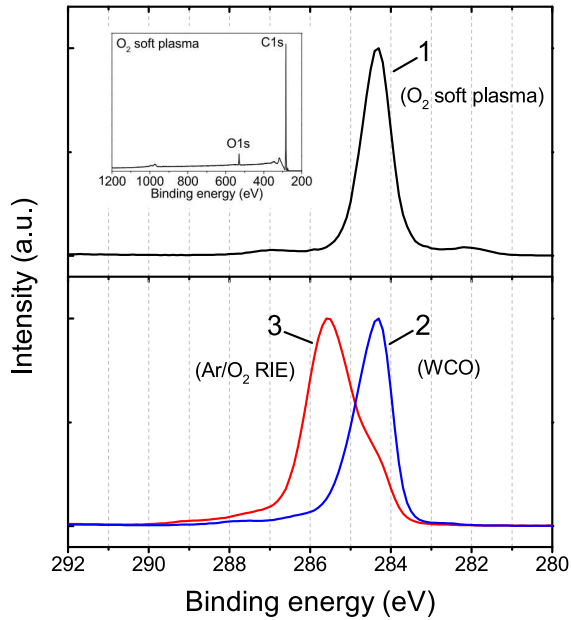}
\caption{\label{XPS} X-ray photoelectron spectroscopy measurements; the different curves represent the oxygen soft plasma-treated sample (curve 1) and the reference sample containing a WCO (curve 2) and Ar/O${}_2$ RIE (curve 3) treated regions. The intensity of the peaks were normalized by the maximum value of the signal.}
\end{figure}

To analyze the effect of the oxygen soft plasma in the properties of the described surface areas, confocal photoluminescence (PL) microscopy (home-build setup with a $532$ nm wavelength excitation laser), AFM and X-ray photoelectron spectroscopy (XPS) measurements were performed. The XPS photoemmision spectra were acquired using a AXIS ULTRA (Kratos Analytical Ltd.) spectrometer equipped with a monochromatized Al K\textit{$\alpha$} (1486.6 eV) radiation source. The binding energy scale was calibrated by means of Ag and Au reference samples. High resolution spectra of the C$1$s core levels detected normal to the sample surface are shown in figure \ref{XPS}. There, the spectrum after the oxygen soft plasma (curve 1) is compared to those from a reference diamond sample. The spectral lines related to WCO and the previously described Ar/O${}_2$ RIE plasma process are shown in curves 2 and 3, respectively. 

The Ar/O${}_2$ RIE treatment has been reported to induce detectable damages to the diamond sub-surface layers\cite{ArO2Denisenko}. In the presented confocal PL microscopy measurement (figure \ref{ICPfigure}(b)), these damages are revealed as a slightly increased luminescence background. The lower part of this PL scan had an additional oxygen soft plasma treatment for 1 minute with $180$ W of ICP power. As it can be seen, this post-treatment was enough to eliminate the background related to the Ar/O${}_2$ RIE process, showing a clear fluorescence contrast. Furthermore, related AFM measurements are presented in figure \ref{ICPfigure} (c). Threating the as-polished and the previously Ar/O${}_2$ RIE etched regions simultaneously with the oxygen soft plasma, approximately $1-2$ nm more in the etching depth (represented by $\Delta$h) could be observed in the Ar/O${}_2$ RIE etched region. Longer exposure times and also different ICP powers did not show significant increase in this difference. This is attributed to the faster initial removal of the RIE-damaged diamond layers. In addition, the corresponding XPS spectrum to the Ar/O${}_2$ RIE plasma process is presented in figure \ref{XPS}, curve 3. It shows a broad peak shifted by approximately $+1.2$ eV with respect to the sp${}^3$ bulk component at $284.3$ eV. This is associated to an amorphous phase of carbon ($\alpha$-sp${}^3$) of a few nanometers in thickness\cite{PlasmaDamXPS}. Importantly, such signature for RIE-induced damages is absent in the C1s core level spectrum from the surface treated with the oxygen soft plasma process, as presented in figure \ref{XPS}, curve 1. Hereafter, the results by confocal PL microscopy, AFM and XPS techniques indicates the complete removal of the amorphous RIE-damaged diamond sub-surface layers by the oxygen soft plasma. Moreover, the extracted thickness of $1 - 2$ nm is in good agreement with the literature\cite{PlasmaDamXPS}. 

Besides, the spectrum showed in figure \ref{XPS}, curve 1 is shown to be very similar to the one related to WCO (curve 2). The dominance of the sp${}^3$ bulk component is observed together with small peaks shifted to higher binding energies, known to be related to carbon-oxygen functional groups on the diamond surface\cite{XPSYagi}. Both oxygen soft plasma- and WCO-treatments show similar intensities of the O$1$s core level spectra (see the inset in figure \ref{XPS}), indicating a full coverage of the surface with oxygen species.

In addition to that, it is also known that new NV centers are formed by post-plasma high temperature annealing\cite{OxPlasmaKim}. In contrast to the surface treated with the Ar/O${}_2$ RIE plasma process, this effect was not observed in the present experiment with an additional oxygen soft plasma treatment before the annealing procedure. This supports further the previously mentioned results, confirming the removal of the sub-surfaces defective layers that would contain plasma-induced vacancies.     

\begin{figure}[b]
\includegraphics[width=\columnwidth]{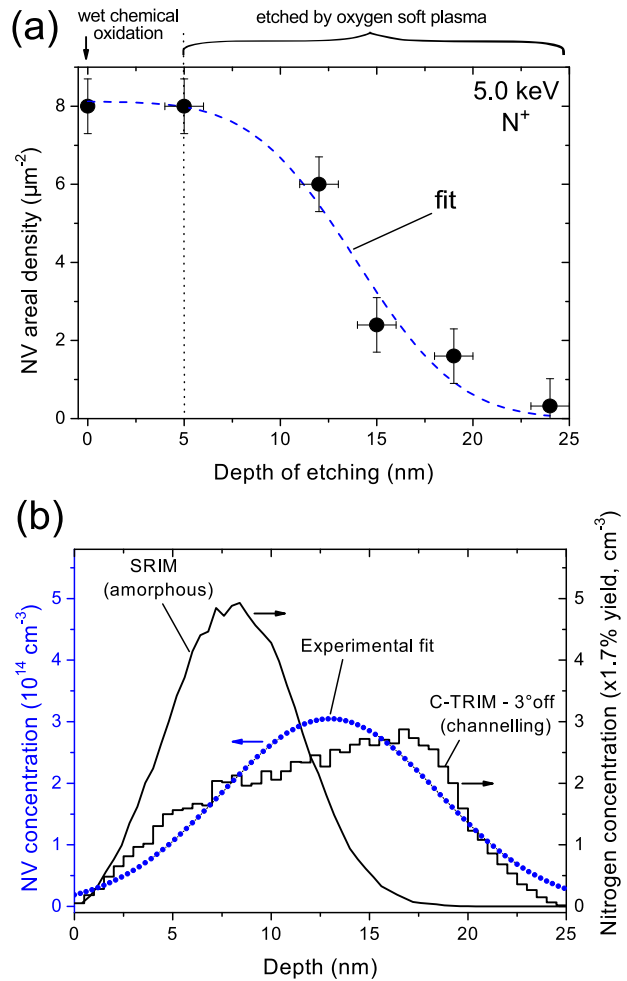}
\caption{\label{figure3} (a): experimental values for the areal density of NV centers by $5.0$ keV of implantation energy vs. the etching depth. The fit corresponds to a Gaussian complementary error function (b): The experimentally fitted depth profile of NV centers is plotted together with results of corresponding SRIM and CTRIM simulations.}
\end{figure}      

Near-surface NV centers can be exquisite tools to detect plasma-induced defects since they are extremely sensitive to surface modifications\cite{OxPlasmaKim,NLplasma,VertDist}. Therefore, to validate the oxygen soft plasma process, the depth profile of NV centers created close to the diamond surface by $5.0$ keV of nitrogen implantation energy with a dose of $4 \times 10^{10}$ cm${}^{-2}$ was analyzed. After the implantation the sample was consequently submitted to high temperature annealing at approximately $950\degree$C in high vacuum ($<10^{-6}$ mbar) for two hours and the mentioned WCO treatment. The chosen annealing temperature is known to minimize the density of paramagnetic defects such as vacancy complexes\cite{Annealing1000}. 

The evaluation consisted of sequential etching steps using the oxygen soft plasma process followed by confocal PL microscopy measurements to extract the areal density of NV centers. Each etching step comprised 1 minute of RIE plasma - aiming low etching rate and homogeneity - followed by 1 minute of the ICP with $180$ W power - aiming the removal of the RIE-induced surface damages (steps referred to the oxygen soft plasma recipe). The experimental areal density of NV centers versus the etching depth is plotted in figure \ref{figure3}(a). For simplicity, the experimental data was fitted to a Gaussian complementary error function. In figure \ref{figure3}(b) the experimentally fitted profile is compared to SRIM\cite{SRIM} and CTRIM\cite{CTRIM} simulations, which consider an amorphous material and ion channeling in a crystalline lattice, respectively. A good agreement was found to the CTRIM simulated profile for a $[100]$-oriented diamond surface with a $3\degree$-off implantation angle, which is in the frame of the accuracy specified for surface polishing and the accuracy of the implantation process. Thus, the presence of \textit{ion channeling} could be resolved even for a low energy of implantation, as also predicted by theoretical investigations using molecular dynamics simulations\cite{DenisMD}. Besides, the yield (conversion from implanted nitrogen atoms to NV centers) was found to be $1.7 \pm 0.3 \%$, a typical value for the used energy of implantation\cite{pezzagnaYield}. It should be highlighted that the point related to the first etching step at $5$ nm (vertical doted line in figure \ref{figure3}(a)) does not show rapid quenching in the density of NV centers in comparison to the initial value (WCO). This contrasts to other plasma etching processes reported in literature\cite{OxPlasmaKim,NLplasma}. Thereby one could ascertain that the developed process preserves the optical properties of shallow NV centers.

The influence of the oxygen soft plasma on the spin coherence characteristics (T${}_2$) of shallow NV centers is of great importance and will be discussed below. To avoid any undesired effects due to polishing-induced defects\cite{PolishingVolpe} or ${}^{13}$C spin bath noise\cite{C13bath}, this study was conducted on an as-grown surface of a ${}^{12}$C isotopically purified ($99.999  \%$) diamond CVD layer with more than $50$ $\mu$m in thickness, overgrown on a natural abundance, single-crystal $[100]$-oriented electronic grade CVD diamond substrate (Element Six).

\begin{figure}
\includegraphics[width=\columnwidth]{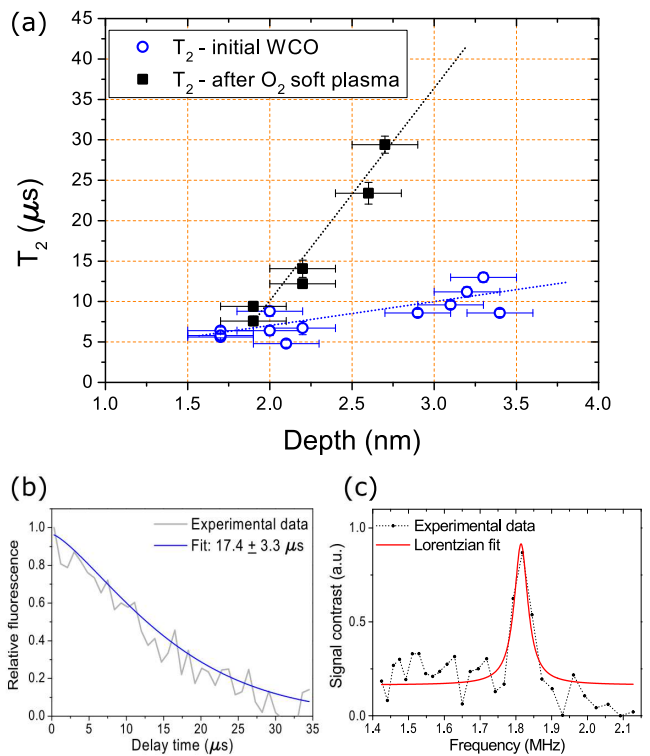}
\caption{\label{T2asgrown} (a): Spin coherence times T${}_2$ (Hahn echo) vs. depth of single NV centers evaluated by the NMR signal generated by protons on the diamond surface. The oxygen soft plasma process is performed and the post-treatment values are compared to the initial ones. (b): an example of measured T${}_2$; no signal originated from ${}^{13}$C spins could be observed. (c): an example of the contrast seen in the fluorescence of the NV center where the frequency of oscillation from protons on the diamond surface is expected.}
\end{figure}

The sample was first implanted with nitrogen ions at $2.5$ keV of energy with a dose of $7 \times 10^{9}$ cm${}^{-2}$ followed by high temperature annealing and WCO as described before. Single NV centers were addressed by confocal microscopy technique. Coherence times were measured by means of optically detected magnetic resonance (ODMR) method using a Hahn echo sequence scheme, which allows us to probe the coherence between the ground state $m_s=\mid 0>$ (bright state) and $m_s=\mid \pm 1>$ (dark state) of the NV center electronic spin. A magnetic field of approximately $420$ Gauss was aligned parallel to the NV axis. Further on, nuclear magnetic resonance (NMR) signal originating from protons present in the immersion oil on the diamond surface was used to calibrate the depth of individual NV centers by means of a XY$8-16$ sequence scheme, as described in reference \cite{staudacher}. The related results are summarized in figure \ref{T2asgrown}.

After treating the initial surface with the oxygen soft plasma process for $2$ minutes with $180$ W of ICP power (corresponding to $2 - 3$ nm of etching depth), NV centers located a few nanometers ($<4$ nm) from the diamond surface could still be observed (figure \ref{T2asgrown}(a)). This supports further the assumption of preservation of the optical properties of shallow NV centers. Likewise, the values of coherence times showed an improvement up to $\sim3$ times, yielding an NMR signal up to $B_{RMS} = 3.1$ $\mu${T}, especially for NV centers located $< 2.5$ nm below the surface. The reason for such improvement is not clear yet, but it can be associated to modifications of the electronic configuration of the diamond surface, which would be affected now by the oxygen soft plasma. Indeed, fluctuations of surface charges are believed to be a significant source of decoherence due to electric field noise\cite{ElecNoise}, meaning that the developed oxygen soft plasma process may increase the surface charge stability. Further studies must be performed to mitigate this effect.

To summarize, an alternative method for etching diamond using an oxygen ICP process was presented. Possessing qualities such as nanometric-precise etching rate, high reproducibility and selective etching, this plasma process was demonstrated to be a promising technique to bring NV centers closer to the diamond surface while preserving their optical and spin characteristics. The presented technique could be used in the future to precisely control the depth of created NV centers by a variety of techniques such as ion and electron irradiation. In addition, being able to profile the depth distribution of NV centers with such precision, one would gain important information about the vacancy diffusion process regarding the creation of color centers in diamond.


F. F. de Oliveira acknowledges the financial support by CNPq project number $204246/2013-0$. J. W. acknowledges the support by the EU via ERC grant SQUTEC and IP DIADEMS as well as the DFG. 

\bibliography{FFOliveira}

\begin{thebibliography}{25}%
\makeatletter
\providecommand \@ifxundefined [1]{%
 \@ifx{#1\undefined}
}%
\providecommand \@ifnum [1]{%
 \ifnum #1\expandafter \@firstoftwo
 \else \expandafter \@secondoftwo
 \fi
}%
\providecommand \@ifx [1]{%
 \ifx #1\expandafter \@firstoftwo
 \else \expandafter \@secondoftwo
 \fi
}%
\providecommand \natexlab [1]{#1}%
\providecommand \enquote  [1]{``#1''}%
\providecommand \bibnamefont  [1]{#1}%
\providecommand \bibfnamefont [1]{#1}%
\providecommand \citenamefont [1]{#1}%
\providecommand \href@noop [0]{\@secondoftwo}%
\providecommand \href [0]{\begingroup \@sanitize@url \@href}%
\providecommand \@href[1]{\@@startlink{#1}\@@href}%
\providecommand \@@href[1]{\endgroup#1\@@endlink}%
\providecommand \@sanitize@url [0]{\catcode `\\12\catcode `\$12\catcode
  `\&12\catcode `\#12\catcode `\^12\catcode `\_12\catcode `\%12\relax}%
\providecommand \@@startlink[1]{}%
\providecommand \@@endlink[0]{}%
\providecommand \url  [0]{\begingroup\@sanitize@url \@url }%
\providecommand \@url [1]{\endgroup\@href {#1}{\urlprefix }}%
\providecommand \urlprefix  [0]{URL }%
\providecommand \Eprint [0]{\href }%
\providecommand \doibase [0]{http://dx.doi.org/}%
\providecommand \selectlanguage [0]{\@gobble}%
\providecommand \bibinfo  [0]{\@secondoftwo}%
\providecommand \bibfield  [0]{\@secondoftwo}%
\providecommand \translation [1]{[#1]}%
\providecommand \BibitemOpen [0]{}%
\providecommand \bibitemStop [0]{}%
\providecommand \bibitemNoStop [0]{.\EOS\space}%
\providecommand \EOS [0]{\spacefactor3000\relax}%
\providecommand \BibitemShut  [1]{\csname bibitem#1\endcsname}%
\let\auto@bib@innerbib\@empty
\bibitem [{\citenamefont {Jelezko}\ and\ \citenamefont
  {Wrachtrup}(2006)}]{reviewJWrachtrup}%
  \BibitemOpen
  \bibfield  {author} {\bibinfo {author} {\bibfnamefont {F.}~\bibnamefont
  {Jelezko}}\ and\ \bibinfo {author} {\bibfnamefont {J.}~\bibnamefont
  {Wrachtrup}},\ }\href {\doibase 10.1002/pssa.200671403} {\bibfield  {journal}
  {\bibinfo  {journal} {phys. stat. sol. (a)}\ }\textbf {\bibinfo {volume}
  {203}},\ \bibinfo {pages} {3207} (\bibinfo {year} {2006})}\BibitemShut
  {NoStop}%
\bibitem [{\citenamefont {Jahnke}\ \emph {et~al.}(2012)\citenamefont {Jahnke},
  \citenamefont {Naydenov}, \citenamefont {Teraji}, \citenamefont {Koizumi},
  \citenamefont {Umeda}, \citenamefont {Isoya},\ and\ \citenamefont
  {Jelezko}}]{JahnkeC12}%
  \BibitemOpen
  \bibfield  {author} {\bibinfo {author} {\bibfnamefont {K.~D.}\ \bibnamefont
  {Jahnke}}, \bibinfo {author} {\bibfnamefont {B.}~\bibnamefont {Naydenov}},
  \bibinfo {author} {\bibfnamefont {T.}~\bibnamefont {Teraji}}, \bibinfo
  {author} {\bibfnamefont {S.}~\bibnamefont {Koizumi}}, \bibinfo {author}
  {\bibfnamefont {T.}~\bibnamefont {Umeda}}, \bibinfo {author} {\bibfnamefont
  {J.}~\bibnamefont {Isoya}}, \ and\ \bibinfo {author} {\bibfnamefont
  {F.}~\bibnamefont {Jelezko}},\ }\href {\doibase
  http://dx.doi.org/10.1063/1.4731778} {\bibfield  {journal} {\bibinfo
  {journal} {Applied Physics Letters}\ }\textbf {\bibinfo {volume} {101}},\
  \bibinfo {pages} {012405} (\bibinfo {year} {2012})}\BibitemShut {NoStop}%
\bibitem [{\citenamefont {Balasubramanian}\ \emph {et~al.}(2009)\citenamefont
  {Balasubramanian}, \citenamefont {Neumann}, \citenamefont {Twitchen},
  \citenamefont {Markham}, \citenamefont {Kolesov}, \citenamefont {Mizuochi},
  \citenamefont {Isoya}, \citenamefont {Achard}, \citenamefont {Beck},
  \citenamefont {Tissler}, \citenamefont {Jacques}, \citenamefont {Hemmer},
  \citenamefont {Jelezko},\ and\ \citenamefont {Wrachtrup}}]{GopanTHz}%
  \BibitemOpen
  \bibfield  {author} {\bibinfo {author} {\bibfnamefont {G.}~\bibnamefont
  {Balasubramanian}}, \bibinfo {author} {\bibfnamefont {P.}~\bibnamefont
  {Neumann}}, \bibinfo {author} {\bibfnamefont {D.}~\bibnamefont {Twitchen}},
  \bibinfo {author} {\bibfnamefont {M.}~\bibnamefont {Markham}}, \bibinfo
  {author} {\bibfnamefont {R.}~\bibnamefont {Kolesov}}, \bibinfo {author}
  {\bibfnamefont {N.}~\bibnamefont {Mizuochi}}, \bibinfo {author}
  {\bibfnamefont {J.}~\bibnamefont {Isoya}}, \bibinfo {author} {\bibfnamefont
  {J.}~\bibnamefont {Achard}}, \bibinfo {author} {\bibfnamefont
  {J.}~\bibnamefont {Beck}}, \bibinfo {author} {\bibfnamefont {J.}~\bibnamefont
  {Tissler}}, \bibinfo {author} {\bibfnamefont {V.}~\bibnamefont {Jacques}},
  \bibinfo {author} {\bibfnamefont {P.~R.}\ \bibnamefont {Hemmer}}, \bibinfo
  {author} {\bibfnamefont {F.}~\bibnamefont {Jelezko}}, \ and\ \bibinfo
  {author} {\bibfnamefont {J.}~\bibnamefont {Wrachtrup}},\ }\href {\doibase
  10.1038/nmat2420} {\bibfield  {journal} {\bibinfo  {journal} {Nature
  Materials}\ }\textbf {\bibinfo {volume} {8}},\ \bibinfo {pages} {383}
  (\bibinfo {year} {2009})}\BibitemShut {NoStop}%
\bibitem [{\citenamefont {Staudacher}\ \emph {et~al.}(2013)\citenamefont
  {Staudacher}, \citenamefont {Shi}, \citenamefont {Pezzagna}, \citenamefont
  {Meijer}, \citenamefont {Du}, \citenamefont {Meriles}, \citenamefont
  {Reinhard},\ and\ \citenamefont {Wrachtrup}}]{staudacher}%
  \BibitemOpen
  \bibfield  {author} {\bibinfo {author} {\bibfnamefont {T.}~\bibnamefont
  {Staudacher}}, \bibinfo {author} {\bibfnamefont {F.}~\bibnamefont {Shi}},
  \bibinfo {author} {\bibfnamefont {S.}~\bibnamefont {Pezzagna}}, \bibinfo
  {author} {\bibfnamefont {J.}~\bibnamefont {Meijer}}, \bibinfo {author}
  {\bibfnamefont {J.}~\bibnamefont {Du}}, \bibinfo {author} {\bibfnamefont
  {C.~A.}\ \bibnamefont {Meriles}}, \bibinfo {author} {\bibfnamefont
  {F.}~\bibnamefont {Reinhard}}, \ and\ \bibinfo {author} {\bibfnamefont
  {J.}~\bibnamefont {Wrachtrup}},\ }\href {\doibase 10.1126/science.1231675}
  {\bibfield  {journal} {\bibinfo  {journal} {Science}\ }\textbf {\bibinfo
  {volume} {339}},\ \bibinfo {pages} {561} (\bibinfo {year}
  {2013})}\BibitemShut {NoStop}%
\bibitem [{\citenamefont {Mamin}\ \emph {et~al.}(2013)\citenamefont {Mamin},
  \citenamefont {Kim}, \citenamefont {Sherwood}, \citenamefont {Rettner},
  \citenamefont {Ohno}, \citenamefont {Awschalom},\ and\ \citenamefont
  {Rugar}}]{MaminNMR2013}%
  \BibitemOpen
  \bibfield  {author} {\bibinfo {author} {\bibfnamefont {H.~J.}\ \bibnamefont
  {Mamin}}, \bibinfo {author} {\bibfnamefont {M.}~\bibnamefont {Kim}}, \bibinfo
  {author} {\bibfnamefont {M.~H.}\ \bibnamefont {Sherwood}}, \bibinfo {author}
  {\bibfnamefont {C.~T.}\ \bibnamefont {Rettner}}, \bibinfo {author}
  {\bibfnamefont {K.}~\bibnamefont {Ohno}}, \bibinfo {author} {\bibfnamefont
  {D.~D.}\ \bibnamefont {Awschalom}}, \ and\ \bibinfo {author} {\bibfnamefont
  {D.}~\bibnamefont {Rugar}},\ }\href {\doibase 10.1126/science.1231540}
  {\bibfield  {journal} {\bibinfo  {journal} {Science}\ }\textbf {\bibinfo
  {volume} {339}},\ \bibinfo {pages} {557} (\bibinfo {year}
  {2013})}\BibitemShut {NoStop}%
\bibitem [{\citenamefont {Müller}\ \emph {et~al.}(2014)\citenamefont
  {Müller}, \citenamefont {Kong}, \citenamefont {Cai}, \citenamefont
  {Melentijevic}, \citenamefont {Stacey}, \citenamefont {Markham},
  \citenamefont {Twitchen}, \citenamefont {Isoya}, \citenamefont {Pezzagna},
  \citenamefont {Meijer}, \citenamefont {Du}, \citenamefont {Plenio},
  \citenamefont {Naydenov}, \citenamefont {McGuinness},\ and\ \citenamefont
  {Jelezko}}]{NMRMueller}%
  \BibitemOpen
  \bibfield  {author} {\bibinfo {author} {\bibfnamefont {C.}~\bibnamefont
  {Müller}}, \bibinfo {author} {\bibfnamefont {X.}~\bibnamefont {Kong}},
  \bibinfo {author} {\bibfnamefont {J.-M.}\ \bibnamefont {Cai}}, \bibinfo
  {author} {\bibfnamefont {K.}~\bibnamefont {Melentijevic}}, \bibinfo {author}
  {\bibfnamefont {A.}~\bibnamefont {Stacey}}, \bibinfo {author} {\bibfnamefont
  {M.}~\bibnamefont {Markham}}, \bibinfo {author} {\bibfnamefont
  {D.}~\bibnamefont {Twitchen}}, \bibinfo {author} {\bibfnamefont
  {J.}~\bibnamefont {Isoya}}, \bibinfo {author} {\bibfnamefont
  {S.}~\bibnamefont {Pezzagna}}, \bibinfo {author} {\bibfnamefont
  {J.}~\bibnamefont {Meijer}}, \bibinfo {author} {\bibfnamefont {J.~F.}\
  \bibnamefont {Du}}, \bibinfo {author} {\bibfnamefont {M.~B.}\ \bibnamefont
  {Plenio}}, \bibinfo {author} {\bibfnamefont {B.}~\bibnamefont {Naydenov}},
  \bibinfo {author} {\bibfnamefont {L.~P.}\ \bibnamefont {McGuinness}}, \ and\
  \bibinfo {author} {\bibfnamefont {F.}~\bibnamefont {Jelezko}},\ }\href
  {\doibase 10.1038/ncomms5703} {\bibfield  {journal} {\bibinfo  {journal}
  {Nature Communications}\ }\textbf {\bibinfo {volume} {5}} (\bibinfo {year}
  {2014}),\ 10.1038/ncomms5703}\BibitemShut {NoStop}%
\bibitem [{\citenamefont {Grotz}\ \emph {et~al.}(2011)\citenamefont {Grotz},
  \citenamefont {Beck}, \citenamefont {Neumann}, \citenamefont {Naydenov},
  \citenamefont {Reuter}, \citenamefont {Reinhard}, \citenamefont {Jelezko},
  \citenamefont {Wrachtrup}, \citenamefont {Schweinfurth}, \citenamefont
  {Sarkar},\ and\ \citenamefont {Hemmer}}]{Bernhard}%
  \BibitemOpen
  \bibfield  {author} {\bibinfo {author} {\bibfnamefont {B.}~\bibnamefont
  {Grotz}}, \bibinfo {author} {\bibfnamefont {J.}~\bibnamefont {Beck}},
  \bibinfo {author} {\bibfnamefont {P.}~\bibnamefont {Neumann}}, \bibinfo
  {author} {\bibfnamefont {B.}~\bibnamefont {Naydenov}}, \bibinfo {author}
  {\bibfnamefont {R.}~\bibnamefont {Reuter}}, \bibinfo {author} {\bibfnamefont
  {F.}~\bibnamefont {Reinhard}}, \bibinfo {author} {\bibfnamefont
  {F.}~\bibnamefont {Jelezko}}, \bibinfo {author} {\bibfnamefont
  {J.}~\bibnamefont {Wrachtrup}}, \bibinfo {author} {\bibfnamefont
  {D.}~\bibnamefont {Schweinfurth}}, \bibinfo {author} {\bibfnamefont
  {B.}~\bibnamefont {Sarkar}}, \ and\ \bibinfo {author} {\bibfnamefont
  {P.}~\bibnamefont {Hemmer}},\ }\href {\doibase 10.1088/1367-2630/13/5/055004}
  {\bibfield  {journal} {\bibinfo  {journal} {New Journal of Physics}\ }\textbf
  {\bibinfo {volume} {13}} (\bibinfo {year} {2011}),\
  10.1088/1367-2630/13/5/055004}\BibitemShut {NoStop}%
\bibitem [{\citenamefont {Shi}\ \emph {et~al.}(2015)\citenamefont {Shi},
  \citenamefont {Zhang}, \citenamefont {Wang}, \citenamefont {Sun},
  \citenamefont {Wang}, \citenamefont {Rong}, \citenamefont {Chen},
  \citenamefont {Ju}, \citenamefont {Reinhard}, \citenamefont {Chen},
  \citenamefont {Wrachtrup}, \citenamefont {Wang},\ and\ \citenamefont
  {Du}}]{singleProteinFazhan}%
  \BibitemOpen
  \bibfield  {author} {\bibinfo {author} {\bibfnamefont {F.}~\bibnamefont
  {Shi}}, \bibinfo {author} {\bibfnamefont {Q.}~\bibnamefont {Zhang}}, \bibinfo
  {author} {\bibfnamefont {P.}~\bibnamefont {Wang}}, \bibinfo {author}
  {\bibfnamefont {H.}~\bibnamefont {Sun}}, \bibinfo {author} {\bibfnamefont
  {J.}~\bibnamefont {Wang}}, \bibinfo {author} {\bibfnamefont {X.}~\bibnamefont
  {Rong}}, \bibinfo {author} {\bibfnamefont {M.}~\bibnamefont {Chen}}, \bibinfo
  {author} {\bibfnamefont {C.}~\bibnamefont {Ju}}, \bibinfo {author}
  {\bibfnamefont {F.}~\bibnamefont {Reinhard}}, \bibinfo {author}
  {\bibfnamefont {H.}~\bibnamefont {Chen}}, \bibinfo {author} {\bibfnamefont
  {J.}~\bibnamefont {Wrachtrup}}, \bibinfo {author} {\bibfnamefont
  {J.}~\bibnamefont {Wang}}, \ and\ \bibinfo {author} {\bibfnamefont
  {J.}~\bibnamefont {Du}},\ }\href {\doibase 10.1126/science.aaa2253}
  {\bibfield  {journal} {\bibinfo  {journal} {Science}\ }\textbf {\bibinfo
  {volume} {347}},\ \bibinfo {pages} {1135} (\bibinfo {year}
  {2015})}\BibitemShut {NoStop}%
\bibitem [{\citenamefont {Pezzagna}\ \emph {et~al.}(2010)\citenamefont
  {Pezzagna}, \citenamefont {Naydenov}, \citenamefont {Jelezko}, \citenamefont
  {Wrachtrup},\ and\ \citenamefont {Meijer}}]{pezzagnaYield}%
  \BibitemOpen
  \bibfield  {author} {\bibinfo {author} {\bibfnamefont {S.}~\bibnamefont
  {Pezzagna}}, \bibinfo {author} {\bibfnamefont {B.}~\bibnamefont {Naydenov}},
  \bibinfo {author} {\bibfnamefont {F.}~\bibnamefont {Jelezko}}, \bibinfo
  {author} {\bibfnamefont {J.}~\bibnamefont {Wrachtrup}}, \ and\ \bibinfo
  {author} {\bibfnamefont {J.}~\bibnamefont {Meijer}},\ }\href {\doibase
  10.1088/1367-2630/12/6/065017} {\bibfield  {journal} {\bibinfo  {journal}
  {New Journal of Physics}\ }\textbf {\bibinfo {volume} {12}},\ \bibinfo
  {pages} {065017} (\bibinfo {year} {2010})}\BibitemShut {NoStop}%
\bibitem [{\citenamefont {Ohno}\ \emph {et~al.}(2012)\citenamefont {Ohno},
  \citenamefont {Joseph~Heremans}, \citenamefont {Bassett}, \citenamefont
  {Myers}, \citenamefont {Toyli}, \citenamefont {Bleszynski~Jayich},
  \citenamefont {Palmstrøm},\ and\ \citenamefont
  {Awschalom}}]{dGrownT2_KOhno}%
  \BibitemOpen
  \bibfield  {author} {\bibinfo {author} {\bibfnamefont {K.}~\bibnamefont
  {Ohno}}, \bibinfo {author} {\bibfnamefont {F.}~\bibnamefont
  {Joseph~Heremans}}, \bibinfo {author} {\bibfnamefont {L.~C.}\ \bibnamefont
  {Bassett}}, \bibinfo {author} {\bibfnamefont {B.~A.}\ \bibnamefont {Myers}},
  \bibinfo {author} {\bibfnamefont {D.~M.}\ \bibnamefont {Toyli}}, \bibinfo
  {author} {\bibfnamefont {A.~C.}\ \bibnamefont {Bleszynski~Jayich}}, \bibinfo
  {author} {\bibfnamefont {C.~J.}\ \bibnamefont {Palmstrøm}}, \ and\ \bibinfo
  {author} {\bibfnamefont {D.~D.}\ \bibnamefont {Awschalom}},\ }\href {\doibase
  http://dx.doi.org/10.1063/1.4748280} {\bibfield  {journal} {\bibinfo
  {journal} {Applied Physics Letters}\ }\textbf {\bibinfo {volume} {101}},\
  \bibinfo {pages} {082413} (\bibinfo {year} {2012})}\BibitemShut {NoStop}%
\bibitem [{\citenamefont {Ohno}\ \emph {et~al.}(2014)\citenamefont {Ohno},
  \citenamefont {Joseph~Heremans}, \citenamefont {de~las Casas}, \citenamefont
  {Myers}, \citenamefont {Alemán}, \citenamefont {Bleszynski~Jayich},\ and\
  \citenamefont {Awschalom}}]{C12irr}%
  \BibitemOpen
  \bibfield  {author} {\bibinfo {author} {\bibfnamefont {K.}~\bibnamefont
  {Ohno}}, \bibinfo {author} {\bibfnamefont {F.}~\bibnamefont
  {Joseph~Heremans}}, \bibinfo {author} {\bibfnamefont {C.~F.}\ \bibnamefont
  {de~las Casas}}, \bibinfo {author} {\bibfnamefont {B.~A.}\ \bibnamefont
  {Myers}}, \bibinfo {author} {\bibfnamefont {B.~J.}\ \bibnamefont {Alemán}},
  \bibinfo {author} {\bibfnamefont {A.~C.}\ \bibnamefont {Bleszynski~Jayich}},
  \ and\ \bibinfo {author} {\bibfnamefont {D.~D.}\ \bibnamefont {Awschalom}},\
  }\href {\doibase 10.1063/1.4890613} {\bibfield  {journal} {\bibinfo
  {journal} {Applied Physics Letters}\ }\textbf {\bibinfo {volume} {105}},\
  \bibinfo {eid} {052406} (\bibinfo {year} {2014}),\
  10.1063/1.4890613}\BibitemShut {NoStop}%
\bibitem [{\citenamefont {Loretz}\ \emph {et~al.}(2014)\citenamefont {Loretz},
  \citenamefont {Pezzagna}, \citenamefont {Meijer},\ and\ \citenamefont
  {Degen}}]{LorentzAB}%
  \BibitemOpen
  \bibfield  {author} {\bibinfo {author} {\bibfnamefont {M.}~\bibnamefont
  {Loretz}}, \bibinfo {author} {\bibfnamefont {S.}~\bibnamefont {Pezzagna}},
  \bibinfo {author} {\bibfnamefont {J.}~\bibnamefont {Meijer}}, \ and\ \bibinfo
  {author} {\bibfnamefont {C.~L.}\ \bibnamefont {Degen}},\ }\href {\doibase
  http://dx.doi.org/10.1063/1.4862749} {\bibfield  {journal} {\bibinfo
  {journal} {Applied Physics Letters}\ }\textbf {\bibinfo {volume} {104}},\
  \bibinfo {pages} {033102} (\bibinfo {year} {2014})}\BibitemShut {NoStop}%
\bibitem [{\citenamefont {Kim}\ \emph {et~al.}(2014)\citenamefont {Kim},
  \citenamefont {Mamin}, \citenamefont {Sherwood}, \citenamefont {Rettner},
  \citenamefont {Frommer},\ and\ \citenamefont {Rugar}}]{OxPlasmaKim}%
  \BibitemOpen
  \bibfield  {author} {\bibinfo {author} {\bibfnamefont {M.}~\bibnamefont
  {Kim}}, \bibinfo {author} {\bibfnamefont {H.~J.}\ \bibnamefont {Mamin}},
  \bibinfo {author} {\bibfnamefont {M.~H.}\ \bibnamefont {Sherwood}}, \bibinfo
  {author} {\bibfnamefont {C.~T.}\ \bibnamefont {Rettner}}, \bibinfo {author}
  {\bibfnamefont {J.}~\bibnamefont {Frommer}}, \ and\ \bibinfo {author}
  {\bibfnamefont {D.}~\bibnamefont {Rugar}},\ }\href {\doibase
  http://dx.doi.org/10.1063/1.4891839} {\bibfield  {journal} {\bibinfo
  {journal} {Applied Physics Letters}\ }\textbf {\bibinfo {volume} {105}},\
  \bibinfo {pages} {042406} (\bibinfo {year} {2014})}\BibitemShut {NoStop}%
\bibitem [{\citenamefont {Santori}\ \emph {et~al.}(2009)\citenamefont
  {Santori}, \citenamefont {Barclay}, \citenamefont {Fu},\ and\ \citenamefont
  {Beausoleil}}]{VertDist}%
  \BibitemOpen
  \bibfield  {author} {\bibinfo {author} {\bibfnamefont {C.}~\bibnamefont
  {Santori}}, \bibinfo {author} {\bibfnamefont {P.~E.}\ \bibnamefont
  {Barclay}}, \bibinfo {author} {\bibfnamefont {K.-M.~C.}\ \bibnamefont {Fu}},
  \ and\ \bibinfo {author} {\bibfnamefont {R.~G.}\ \bibnamefont {Beausoleil}},\
  }\href {\doibase http://dx.doi.org/10.1103/PhysRevB.79.125313} {\bibfield
  {journal} {\bibinfo  {journal} {Physical Review B}\ }\textbf {\bibinfo
  {volume} {79}},\ \bibinfo {pages} {125313} (\bibinfo {year}
  {2009})}\BibitemShut {NoStop}%
\bibitem [{\citenamefont {Cui}\ \emph {et~al.}(2015)\citenamefont {Cui},
  \citenamefont {Greenspon}, \citenamefont {Ohno}, \citenamefont {Myers},
  \citenamefont {Jayich}, \citenamefont {Awschalom},\ and\ \citenamefont
  {Hu}}]{NLplasma}%
  \BibitemOpen
  \bibfield  {author} {\bibinfo {author} {\bibfnamefont {S.}~\bibnamefont
  {Cui}}, \bibinfo {author} {\bibfnamefont {A.~S.}\ \bibnamefont {Greenspon}},
  \bibinfo {author} {\bibfnamefont {K.}~\bibnamefont {Ohno}}, \bibinfo {author}
  {\bibfnamefont {B.~A.}\ \bibnamefont {Myers}}, \bibinfo {author}
  {\bibfnamefont {A.~C.~B.}\ \bibnamefont {Jayich}}, \bibinfo {author}
  {\bibfnamefont {D.~D.}\ \bibnamefont {Awschalom}}, \ and\ \bibinfo {author}
  {\bibfnamefont {E.~L.}\ \bibnamefont {Hu}},\ }\href {\doibase
  10.1021/acs.nanolett.5b00457} {\bibfield  {journal} {\bibinfo  {journal}
  {Nano Letters}\ }\textbf {\bibinfo {volume} {15(5)}},\ \bibinfo {pages}
  {2887} (\bibinfo {year} {2015})}\BibitemShut {NoStop}%
\bibitem [{\citenamefont {Kawabata}\ \emph {et~al.}(1988)\citenamefont
  {Kawabata}, \citenamefont {Taniguchi},\ and\ \citenamefont
  {Miyamoto}}]{PlasmaDamXPS}%
  \BibitemOpen
  \bibfield  {author} {\bibinfo {author} {\bibfnamefont {Y.}~\bibnamefont
  {Kawabata}}, \bibinfo {author} {\bibfnamefont {J.}~\bibnamefont {Taniguchi}},
  \ and\ \bibinfo {author} {\bibfnamefont {I.}~\bibnamefont {Miyamoto}},\
  }\href {\doibase 10.1016/j.diamond.2003.09.005} {\bibfield  {journal}
  {\bibinfo  {journal} {Diamond and Related Materials}\ }\textbf {\bibinfo
  {volume} {13}},\ \bibinfo {pages} {93 } (\bibinfo {year} {1988})}\BibitemShut
  {NoStop}%
\bibitem [{\citenamefont {Denisenko}\ \emph {et~al.}(2010)\citenamefont
  {Denisenko}, \citenamefont {Romanyuk}, \citenamefont {Pietzka}, \citenamefont
  {Scharpf},\ and\ \citenamefont {Kohn}}]{ArO2Denisenko}%
  \BibitemOpen
  \bibfield  {author} {\bibinfo {author} {\bibfnamefont {A.}~\bibnamefont
  {Denisenko}}, \bibinfo {author} {\bibfnamefont {A.}~\bibnamefont {Romanyuk}},
  \bibinfo {author} {\bibfnamefont {C.}~\bibnamefont {Pietzka}}, \bibinfo
  {author} {\bibfnamefont {J.}~\bibnamefont {Scharpf}}, \ and\ \bibinfo
  {author} {\bibfnamefont {E.}~\bibnamefont {Kohn}},\ }\href {\doibase
  http://dx.doi.org/10.1063/1.3489986} {\bibfield  {journal} {\bibinfo
  {journal} {Journal of Applied Physics}\ }\textbf {\bibinfo {volume} {108}},\
  \bibinfo {pages} {074901} (\bibinfo {year} {2010})}\BibitemShut {NoStop}%
\bibitem [{\citenamefont {Yagi}\ \emph {et~al.}(1999)\citenamefont {Yagi},
  \citenamefont {Notsu}, \citenamefont {Kondo}, \citenamefont {Tryk},\ and\
  \citenamefont {Fujishima}}]{XPSYagi}%
  \BibitemOpen
  \bibfield  {author} {\bibinfo {author} {\bibfnamefont {I.}~\bibnamefont
  {Yagi}}, \bibinfo {author} {\bibfnamefont {H.}~\bibnamefont {Notsu}},
  \bibinfo {author} {\bibfnamefont {T.}~\bibnamefont {Kondo}}, \bibinfo
  {author} {\bibfnamefont {D.~A.}\ \bibnamefont {Tryk}}, \ and\ \bibinfo
  {author} {\bibfnamefont {A.}~\bibnamefont {Fujishima}},\ }\href {\doibase
  10.1016/S0022-0728(99)00027-3} {\bibfield  {journal} {\bibinfo  {journal}
  {Journal of Electroanalytical Chemistry}\ }\textbf {\bibinfo {volume}
  {473}},\ \bibinfo {pages} {173} (\bibinfo {year} {1999})}\BibitemShut
  {NoStop}%
\bibitem [{\citenamefont {Yamamoto}\ \emph {et~al.}(2013)\citenamefont
  {Yamamoto}, \citenamefont {Umeda}, \citenamefont {Watanabe}, \citenamefont
  {Onoda}, \citenamefont {Markham}, \citenamefont {Twitchen}, \citenamefont
  {Naydenov}, \citenamefont {McGuinness}, \citenamefont {Teraji}, \citenamefont
  {Koizumi}, \citenamefont {Dolde}, \citenamefont {Fedder}, \citenamefont
  {Honert}, \citenamefont {Wrachtrup}, \citenamefont {Ohshima}, \citenamefont
  {Jelezko},\ and\ \citenamefont {Isoya}}]{Annealing1000}%
  \BibitemOpen
  \bibfield  {author} {\bibinfo {author} {\bibfnamefont {T.}~\bibnamefont
  {Yamamoto}}, \bibinfo {author} {\bibfnamefont {T.}~\bibnamefont {Umeda}},
  \bibinfo {author} {\bibfnamefont {K.}~\bibnamefont {Watanabe}}, \bibinfo
  {author} {\bibfnamefont {S.}~\bibnamefont {Onoda}}, \bibinfo {author}
  {\bibfnamefont {M.~L.}\ \bibnamefont {Markham}}, \bibinfo {author}
  {\bibfnamefont {D.~J.}\ \bibnamefont {Twitchen}}, \bibinfo {author}
  {\bibfnamefont {B.}~\bibnamefont {Naydenov}}, \bibinfo {author}
  {\bibfnamefont {L.~P.}\ \bibnamefont {McGuinness}}, \bibinfo {author}
  {\bibfnamefont {T.}~\bibnamefont {Teraji}}, \bibinfo {author} {\bibfnamefont
  {S.}~\bibnamefont {Koizumi}}, \bibinfo {author} {\bibfnamefont
  {F.}~\bibnamefont {Dolde}}, \bibinfo {author} {\bibfnamefont
  {H.}~\bibnamefont {Fedder}}, \bibinfo {author} {\bibfnamefont
  {J.}~\bibnamefont {Honert}}, \bibinfo {author} {\bibfnamefont
  {J.}~\bibnamefont {Wrachtrup}}, \bibinfo {author} {\bibfnamefont
  {T.}~\bibnamefont {Ohshima}}, \bibinfo {author} {\bibfnamefont
  {F.}~\bibnamefont {Jelezko}}, \ and\ \bibinfo {author} {\bibfnamefont
  {J.}~\bibnamefont {Isoya}},\ }\href {\doibase 10.1103/PhysRevB.88.075206}
  {\bibfield  {journal} {\bibinfo  {journal} {Physical Review B}\ }\textbf
  {\bibinfo {volume} {88}},\ \bibinfo {pages} {075206} (\bibinfo {year}
  {2013})}\BibitemShut {NoStop}%
\bibitem [{\citenamefont {Ziegler}\ \emph {et~al.}()\citenamefont {Ziegler},
  \citenamefont {Biersack},\ and\ \citenamefont {Ziegler}}]{SRIM}%
  \BibitemOpen
  \bibfield  {author} {\bibinfo {author} {\bibfnamefont {J.~F.}\ \bibnamefont
  {Ziegler}}, \bibinfo {author} {\bibfnamefont {J.~P.}\ \bibnamefont
  {Biersack}}, \ and\ \bibinfo {author} {\bibfnamefont {M.~D.}\ \bibnamefont
  {Ziegler}},\ }\href {http://www.srim.org/} {\enquote {\bibinfo {title} {Srim,
  the stopping and range of ions in matter},}\ }\BibitemShut {NoStop}%
\bibitem [{\citenamefont {Posselt}\ and\ \citenamefont
  {Biersack}(1992)}]{CTRIM}%
  \BibitemOpen
  \bibfield  {author} {\bibinfo {author} {\bibfnamefont {M.}~\bibnamefont
  {Posselt}}\ and\ \bibinfo {author} {\bibfnamefont {J.~P.}\ \bibnamefont
  {Biersack}},\ }\href {\doibase 10.1016/0168-583X(92)95562-6} {\bibfield
  {journal} {\bibinfo  {journal} {Nucl. Instr. Meth. Phys. Res. Sec. B}\
  }\textbf {\bibinfo {volume} {64}},\ \bibinfo {pages} {706} (\bibinfo {year}
  {1992})}\BibitemShut {NoStop}%
\bibitem [{\citenamefont {Antonov}\ \emph {et~al.}(2014)\citenamefont
  {Antonov}, \citenamefont {Häußermann}, \citenamefont {Aird}, \citenamefont
  {Roth}, \citenamefont {Trebin}, \citenamefont {Müller}, \citenamefont
  {McGuinness}, \citenamefont {Jelezko}, \citenamefont {Yamamoto},
  \citenamefont {Isoya}, \citenamefont {Pezzagna}, \citenamefont {Meijer},\
  and\ \citenamefont {Wrachtrup}}]{DenisMD}%
  \BibitemOpen
  \bibfield  {author} {\bibinfo {author} {\bibfnamefont {D.}~\bibnamefont
  {Antonov}}, \bibinfo {author} {\bibfnamefont {T.}~\bibnamefont
  {Häußermann}}, \bibinfo {author} {\bibfnamefont {A.}~\bibnamefont {Aird}},
  \bibinfo {author} {\bibfnamefont {J.}~\bibnamefont {Roth}}, \bibinfo {author}
  {\bibfnamefont {H.-R.}\ \bibnamefont {Trebin}}, \bibinfo {author}
  {\bibfnamefont {C.}~\bibnamefont {Müller}}, \bibinfo {author} {\bibfnamefont
  {L.}~\bibnamefont {McGuinness}}, \bibinfo {author} {\bibfnamefont
  {F.}~\bibnamefont {Jelezko}}, \bibinfo {author} {\bibfnamefont
  {T.}~\bibnamefont {Yamamoto}}, \bibinfo {author} {\bibfnamefont
  {J.}~\bibnamefont {Isoya}}, \bibinfo {author} {\bibfnamefont
  {S.}~\bibnamefont {Pezzagna}}, \bibinfo {author} {\bibfnamefont
  {J.}~\bibnamefont {Meijer}}, \ and\ \bibinfo {author} {\bibfnamefont
  {J.}~\bibnamefont {Wrachtrup}},\ }\href {\doibase
  http://dx.doi.org/10.1063/1.4860997} {\bibfield  {journal} {\bibinfo
  {journal} {Applied Physics Letters}\ }\textbf {\bibinfo {volume} {104}},\
  \bibinfo {pages} {012105} (\bibinfo {year} {2014})}\BibitemShut {NoStop}%
\bibitem [{\citenamefont {Volpe}\ \emph {et~al.}(2009)\citenamefont {Volpe},
  \citenamefont {Muret}, \citenamefont {Omnes}, \citenamefont {Achard},
  \citenamefont {Silva}, \citenamefont {Brinza},\ and\ \citenamefont
  {Gicquel}}]{PolishingVolpe}%
  \BibitemOpen
  \bibfield  {author} {\bibinfo {author} {\bibfnamefont {P.-N.}\ \bibnamefont
  {Volpe}}, \bibinfo {author} {\bibfnamefont {P.}~\bibnamefont {Muret}},
  \bibinfo {author} {\bibfnamefont {F.}~\bibnamefont {Omnes}}, \bibinfo
  {author} {\bibfnamefont {J.}~\bibnamefont {Achard}}, \bibinfo {author}
  {\bibfnamefont {F.}~\bibnamefont {Silva}}, \bibinfo {author} {\bibfnamefont
  {O.}~\bibnamefont {Brinza}}, \ and\ \bibinfo {author} {\bibfnamefont
  {A.}~\bibnamefont {Gicquel}},\ }\href {\doibase
  http://dx.doi.org/10.1016/j.diamond.2009.04.008} {\bibfield  {journal}
  {\bibinfo  {journal} {Diamond and Related Materials}\ }\textbf {\bibinfo
  {volume} {18}},\ \bibinfo {pages} {1205 } (\bibinfo {year}
  {2009})}\BibitemShut {NoStop}%
\bibitem [{\citenamefont {Markham}\ \emph {et~al.}(2011)\citenamefont
  {Markham}, \citenamefont {Dodson}, \citenamefont {Scarsbrook}, \citenamefont
  {Twitchen}, \citenamefont {Balasubramanian}, \citenamefont {Jelezko},\ and\
  \citenamefont {Wrachtrup}}]{C13bath}%
  \BibitemOpen
  \bibfield  {author} {\bibinfo {author} {\bibfnamefont {M.}~\bibnamefont
  {Markham}}, \bibinfo {author} {\bibfnamefont {J.}~\bibnamefont {Dodson}},
  \bibinfo {author} {\bibfnamefont {G.}~\bibnamefont {Scarsbrook}}, \bibinfo
  {author} {\bibfnamefont {D.}~\bibnamefont {Twitchen}}, \bibinfo {author}
  {\bibfnamefont {G.}~\bibnamefont {Balasubramanian}}, \bibinfo {author}
  {\bibfnamefont {F.}~\bibnamefont {Jelezko}}, \ and\ \bibinfo {author}
  {\bibfnamefont {J.}~\bibnamefont {Wrachtrup}},\ }\href {\doibase
  10.1016/j.diamond.2010.11.016} {\bibfield  {journal} {\bibinfo  {journal}
  {Diamond and Related Materials}\ }\textbf {\bibinfo {volume} {20}},\ \bibinfo
  {pages} {134 } (\bibinfo {year} {2011})}\BibitemShut {NoStop}%
\bibitem [{\citenamefont {{Kim}}\ \emph {et~al.}(2015)\citenamefont {{Kim}},
  \citenamefont {{Mamin}}, \citenamefont {{Sherwood}}, \citenamefont {{Ohno}},
  \citenamefont {{Awschalom}},\ and\ \citenamefont {{Rugar}}}]{ElecNoise}%
  \BibitemOpen
  \bibfield  {author} {\bibinfo {author} {\bibfnamefont {M.}~\bibnamefont
  {{Kim}}}, \bibinfo {author} {\bibfnamefont {H.~J.}\ \bibnamefont {{Mamin}}},
  \bibinfo {author} {\bibfnamefont {M.~H.}\ \bibnamefont {{Sherwood}}},
  \bibinfo {author} {\bibfnamefont {K.}~\bibnamefont {{Ohno}}}, \bibinfo
  {author} {\bibfnamefont {D.~D.}\ \bibnamefont {{Awschalom}}}, \ and\ \bibinfo
  {author} {\bibfnamefont {D.}~\bibnamefont {{Rugar}}},\ }\href@noop {}
  {\bibfield  {journal} {\bibinfo  {journal} {ArXiv e-prints}\ } (\bibinfo
  {year} {2015})},\ \Eprint {http://arxiv.org/abs/1506.00295} {arXiv:1506.00295
  [cond-mat.mes-hall]} \BibitemShut {NoStop}%
\end{thebibliography}%

\end{document}